\documentclass{aa}
\usepackage{graphicx}
\usepackage{txfonts}
\usepackage{hyperref}

\bibpunct{(}{)}{;}{a}{}{,}

\begin{document}

   \title{Evolution of cataclysmic variables under different magnetic braking prescriptions}

   \subtitle{}

   \author{Bingyao Zhou\inst{1}
        \and
        Chunhua Zhu\inst{1}\fnmsep\thanks{Corresponding author: chunhuazhu@sina.cn}
        \and
        Guoliang L{\"u}\inst{1}
        \and
        Sufen Guo\inst{1}
        \and
        Helei Liu\inst{1}
        \and
        Farkhodjon Khamrakulov\inst{2}
        }

   \institute{School of Physical Science and Technology, Xinjiang University, Urumqi 830017, China
        \email{chunhuazhu@sina.cn}
   \and
        Department of Nuclear Physics and Astronomy, Institute of Engineering Physics, Samarkand State University, Samarkand, 140104, Uzbekistan
   }
   \date{Received January 22, 2026} 

 
  \abstract
   {The evolution of cataclysmic variables (CVs) – interacting binaries where a low-mass donor transfers matter to a white dwarf via an accretion disk – is critically controlled by magnetic braking (MB). Significant uncertainties persist regarding how distinct MB formalisms influence CV evolutionary pathways.}
   {We performed systematic simulations of CV evolution under five MB prescriptions using the \texttt{MESA} code: the classical Skumanich law and the Matt, Reiners \& Mohanty (RM12), intermediate, and convection-boosted formalisms. Primary objectives included investigating their impact on orbital period distributions, mass-transfer rates, donor star evolution, and period gap characteristics.}
   {Evolutionary sequences were computed across all MB frameworks. We analyzed their effects on key observables: orbital period evolution, accretion rates, and period gap morphology.}
   {Magnetic braking prescription selection fundamentally determines whether CV systems develop the characteristic period gap. The intermediate prescription provides optimal consistency with observations of nonmagnetic CVs, simultaneously reproducing the gap location and donor properties. Strong braking models (e.g., Skumanich) produce clear detachment phases, while self-consistent regulation models (Matt12 and RM12) maintain weak angular momentum loss and fail to form a gap, making them more prone to magnetic CVs.}
   {The presence or absence of the period gap is primarily governed by the strength and behavior of MB before the donor becomes fully convective. Future studies must further incorporate the regulatory effects of magnetic fields on donor structure to accurately predict the period distribution characteristics of magnetic CVs.}

   \keywords{stars: evolution --
          stars: magnetic field --
          stars: mass-loss --
          stars: novae, cataclysmic variables --
          stars: white dwarfs}

   \maketitle
   \nolinenumbers

\section{Introduction}
Cataclysmic variables (CVs) are close binary systems consisting of a white dwarf (WD) accreting matter from a low-mass companion star that fills its Roche lobe, transferring material through the L1 Lagrangian point. These systems serve as essential laboratories for studying close binary evolution, accretion physics, and stellar magnetic activity \citep{Warner1995, Cannizzo1988, Knigge2011}. One of the most intriguing features in the CV population is the observed ``period gap,'' a significant deficit of systems with orbital periods between approximately 2 and 3 hours. This gap has been interpreted as evidence of a temporary cessation of mass transfer, likely caused by disrupted magnetic braking (MB) when the secondary becomes fully convective \citep{Rappaport1983, spruit1983}.

Magnetic braking plays a crucial role in the angular momentum loss (AML) mechanism driving CV evolution. The process occurs when ionized particles in the stellar wind follow magnetic field lines, effectively creating a lever arm that extracts angular momentum from the rotating star \citep{Matt2008}. When MB ceases and AML decreases, the donor, which had previously expanded beyond its thermal-equilibrium radius, thermally relaxes and contracts within its Roche lobe. This temporarily halts mass transfer until gravitational radiation drives the system's components back into contact with each other, allowing accretion to resume \citep{Howell2001}. The efficiency of this MB mechanism fundamentally determines the evolutionary pathways of CVs, including the location and width of the period gap \citep{Knigge2011}.

Several prescriptions for MB have been proposed over the decades. The classical Skumanich law \citep{Skumanich1972}, which establishes the relationship between the stellar rotation rate and age ($\Omega \propto t^{-1/2}$), forms the basis for standard MB prescriptions in CV evolution models \citep{Verbunt1981}. However, recent observations of stellar rotation periods in open clusters challenge the classical Skumanich relation. Studies indicate that this relation does not universally apply across all stellar masses and rotational regimes \citep{Van2016, Metcalfe2016, Hall2021}. Notably, low-mass stars in intermediate-age open clusters (e.g., NGC 6811 and Ruprecht 147) exhibit a pronounced departure from expected spin-down trends, characterized by rotational evolution plateaus \citep{Curtis2019, Curtis2020, Santos2025}. This anomalous behavior suggests complex AML mechanisms, motivating sophisticated models that incorporate dependences on stellar mass, radius, magnetic field strength, and rotation rate.

Among these improved models, \citet{Matt2012} proposed a formula based on magnetohydrodynamic stellar wind simulations, accounting for dependences of AML on the magnetic field strength (particularly di-polar fields) and stellar rotation rate. This model precisely computes the interaction between stellar winds and rotating magnetic fields, revealing the critical roles of rotation and donor star magnetic field strength in AML. Furthermore, \citet{Reiners2012} introduced a radius-dependent model that emphasizes the impact of changes in magnetic topology on the evolution of angular momentum in low-mass stars, especially during the transition to full convection. According to this model, variations in stellar radius play a crucial role in MB for low-mass stars, particularly as stars become fully convective and undergo significant changes in magnetic field geometry and strength, thereby affecting AML processes. Beyond these models, \citet{Van2019} developed novel modifications to traditional prescriptions through adjusted power-law indices, creating ``intermediate'' and ``boosted'' MB models. By tuning these power-law exponents, their formulation incorporates not only the rotation rate but also additional factors such as wind mass loss and convective timescales, elements that were not adequately addressed in the original \citet{Skumanich1972} relation. Although extensively tested for low-mass X-ray binaries (LMXBs; \citealt{Deng2021, Echeveste2024}), comprehensive comparative studies of the impact of these prescriptions on CV evolution remain scarce. Previous investigations of CV evolution typically employed conventional braking models, with minimal systematic comparisons between fundamentally different MB formulations.

Cataclysmic variables can be broadly divided into nonmagnetic and magnetic systems. The magnetic CVs are further grouped into intermediate polars (IPs) and polars \citep{Mauche1997, Cropper1990}. In polars, strong magnetic fields ($\sim 10^7$--$10^8$ G) synchronize the rotation of the WD with the binary orbit, suppressing accretion disk formation and channeling material directly along field lines onto the WD. In contrast, in IPs, weaker magnetic fields ($\sim 10^6$--$10^7$ G) fail to achieve synchronization, allowing most systems to form truncated accretion disks \citep{Worpel2020, Belloni2020}. Magnetic fields likely play crucial roles in CV evolution by modulating both the mechanisms and the rates of angular momentum transport. Systems with different field strengths exhibit distinct physical configurations and consequently follow divergent evolutionary pathways.

Recent large-scale surveys --- such as the Sloan Digital Sky Survey and surveys conducted with the Large Sky Area Multi-Object Fiber Spectroscopic Telescope and the \textit{Gaia} space telescope ---have dramatically expanded the known populations of both nonmagnetic and magnetic CVs. They have provided unprecedented statistical samples for theoretical modeling \citep{Inight2023,Inight2025, Sun2021, Munoz2024, Pala2020}. \citep{McAllister2019,van2025}. As MB represents a primary AML mechanism in CV evolution, determining which MB prescription can reconcile these observational constraints constitutes the central focus of our current investigation.

We conducted a comparative study of five distinct MB prescriptions and their impact on CV evolution: the classical \cite{Skumanich1972} model, the  \cite{Matt2012} formulation, the \cite{Reiners2012} model, an intermediate prescription \citep{Van2019}, and a boosted prescription, in which the power-law exponents differ from the intermediate case. We used the stellar evolution code \texttt{MESA} (Modules for Experiments in Stellar Astrophysics) to simulate CV evolution under these different prescriptions and analyze their implications on observable properties. Particular attention is given to the orbital period distribution, mass-transfer rates, and the mass--radius relationship of donor stars, which are crucial for testing theoretical models against observations. Furthermore, we explored a wide parameter space by varying the initial masses of both the WD and its companion, ensuring robust coverage of possible CV configurations. This comprehensive treatment allowed us to assess the sensitivity of CV evolution to different AML prescriptions in a consistent and unified framework.

In Sect. 2 we provide a stellar evolution code, the configuration of our binary model, and the mathematical formulations of the five MB prescriptions. Section 3 presents the results of our simulations, focusing on the evolution of the orbital period, mass-transfer rate, and donor star characteristics. We compare the theoretical predictions with observational data and investigate the effects of the magnetic field on mass loss and the period gap. In Sect. 4 we comprehensively discuss and analyze magnetic-field-dependent models in comparison with statistical observational data. Finally, Sect. 5 summarizes our main conclusions and suggests directions for future research.
\section{Evolution code and binary model}
\subsection{The stellar evolution code}
For binary evolutionary calculations, we employed the \texttt{MESA} code, version 15140 \citep{Paxton2011,Paxton2013,Paxton2015,Paxton2018,Paxton2019}. \texttt{MESA} is a state-of-the-art, open-source stellar evolution code that incorporates detailed physics essential for modeling the stellar structure and evolution. The simulated binaries initially consist of a main sequence donor star (of mass $M_{\mathrm{d}}$) and a WD (of mass $M_{\mathrm{WD}}$). The formation and evolution of such WD + main-sequence star systems have been extensively discussed in the literature (e.g., \citealt{Kool1992, Lu2006, Nelemans2016, Zhu2023, Gao2024, Wang2025, Lu2020, Zhu2021}). We configured \texttt{MESA} with solar metallicity (Z = 0.02) and mixing length parameter $\alpha = 2.0$ using the Henyey version of the mixing-length theory (MLT) \citep{Henyey1965}. The equation of state combines FreeEOS \citep{Irwin2012}, OPAL\_SCVH \citep{Rogers2002,Saumon1995}, and Skye \citep{Jermyn2021} components.

In our work, we treated the WD primary as a point mass following the approach of \citet{Ritter1988}. This approximation is justified because the radius of the WD is typically two orders of magnitude smaller than the orbital separation in CV systems, and its internal structure has minimal impact on the orbital evolution compared to the effects of mass-transfer and AML. The point mass approximation significantly reduces computational complexity while maintaining the essential physics governing CV evolution.

For the donor star, we evolved a detailed stellar structure model that dynamically responds to mass loss via Roche-lobe overflow. The mass-transfer rate was computed using the prescription provided by \citet{Ritter1988}, which offers a smooth and physically motivated transition between detached systems and steady-state Roche-lobe overflow:
\begin{equation}
\dot{M}_{\mathrm{1}} = -\dot{M}_0 \exp\left(\frac{R_1 - R_{\mathrm{L},1}}{H_{\mathrm{P}}}\right),
\end{equation}
where $\dot{M}_{\mathrm{1}}$ is the mass-transfer rate from the donor, $R_1$ is the donor radius, $R_{\mathrm{L},1}$ is the Roche-lobe radius of the donor, and $H_{\mathrm{P}}$ is the pressure scale height at the photosphere of the donor. The parameter $\dot{M}_0$ is a characteristic mass-transfer rate, expressed as
\[
\dot{M}_0 = \frac{1}{\sqrt{e}} \, \rho_{\text{ph}} \, c_s \, S,
\]
where $\rho_{\text{ph}}$ denotes the density at the donor's photosphere, $c_s$ is the sound speed, and $S$ represents the effective cross-sectional area of the flow through the inner Lagrangian point. This exponential prescription accounts for mass transfer through the L1 point even when the donor slightly underfills its Roche lobe, due to its extended atmosphere. The Roche-lobe radius of the donor is approximated using the fitting formula from \citet{Eggleton1983}:
\begin{equation}
\frac{R_{\mathrm{L},1}}{a} = \frac{0.49 q^{2/3}}{0.6 q^{2/3} + \ln(1 + q^{1/3})},
\end{equation}where $a$ is the orbital separation and $q = M_1 / M_2$ is the mass ratio of the donor to the accretor. The orbital angular momentum of the binary system is given by
\begin{equation}
J = M_1 M_2 \sqrt{\frac{G a}{M_1 + M_2}},
\end{equation}
where $G$ is the gravitational constant. In our calculations, we assumed that the mass transfer is fully nonconservative. All material transferred to the WD is expelled from the system via isotropic re-emission from the vicinity of the WD, so that the WD mass remains approximately constant during the evolution. Following the isotropic re-emission prescription of \citet{Soberman1997}, the expelled material carries away the specific orbital angular momentum of the accretor. Under this prescription, the specific angular momentum carried away per unit mass lost is
\begin{equation}
\mathrm{j}_{\mathrm{loss}} = \left( \frac{M_{\mathrm{d}}}{M_{\mathrm{tot}}} \right)^{2} a^{2} \Omega ,
\end{equation}
This expression was obtained by evaluating the orbital specific angular momentum of the WD in the isotropic re-emission formalism of \citet{Soberman1997}.
The rate of AML due to systemic mass-transfer is
\begin{equation}
\left( \frac{\dot{J}}{J} \right)_{\mathrm{ML}} 
= \frac{\dot{M}_{\mathrm{loss}}}{M_{\mathrm{tot}}} \cdot j_{\mathrm{loss}},
\end{equation}
where $\dot{M}_{\mathrm{loss}}$ is the total mass lost from the system, 
$M_{\mathrm{tot}} = M_1 + M_2$ is the total mass, and $j_{\mathrm{loss}}$ is the specific 
angular momentum carried away per unit mass lost.
We also included AML due to gravitational wave radiation, using the 
standard quadrupole formula \citep{Landau1975}:
\begin{equation}
\left( \frac{\dot{J}}{J} \right)_{\mathrm{GR}} 
= -\frac{32 G^3}{5 c^5} \cdot 
\frac{M_1 M_2 (M_1 + M_2)}{a^4},
\end{equation}
where $c$ is the speed of light. This mechanism becomes especially important for compact 
binaries with orbital periods shorter than about two hours, and dominates the orbital 
evolution of CVs below the period gap \citep{Paczynski1967, Faulkner1971}. 
The total AML rate is
\begin{equation}
\frac{\dot{J}}{J} 
= \left( \frac{\dot{J}}{J} \right)_{\mathrm{ML}}
+ \left( \frac{\dot{J}}{J} \right)_{\mathrm{GR}}
+ \left( \frac{\dot{J}}{J} \right)_{\mathrm{MB}},
\end{equation}
The final term accounts for AML due to MB. 
We explore several models for MB in the next section.

\subsection{Magnetic braking prescriptions}
In this study, we implemented and compared five distinct MB prescriptions, each representing different theoretical approaches to modeling AML via magnetized stellar winds. These models differ primarily in their dependence on stellar parameters such as radius, mass, rotation rate, and magnetic field strength.

\subsubsection{Skumanich model}
The classical Skumanich model is based on the empirical relation between stellar rotation and age derived from observations of solar-type stars in open clusters \citep{Skumanich1972}. In the context of binary evolution, this was formulated by \citet{Verbunt1981} and later expressed in a more calibrated form as
\begin{equation}
\dot{J}_{\text{mb,Sk}} = -3.8 \times 10^{-30} M_1 R_\odot^4 \left( \frac{R_1}{R_\odot} \right)^{\gamma_{\text{mb}}} \Omega^3 \text{ dyne cm,}
\end{equation}where $R_1$ and $M_1
$ are the radius and mass of the donor star, $\Omega$ is the angular velocity, and $\gamma_{\text{mb}}$ is a parameter that characterizes the dependence on stellar radius (commonly taken as $\gamma_{\text{mb}} = 4$). In tidally locked binaries, $\Omega$ equals the orbital angular velocity.

\begin{figure*}[ht!]
  \centering
  \includegraphics[width=0.48\textwidth]{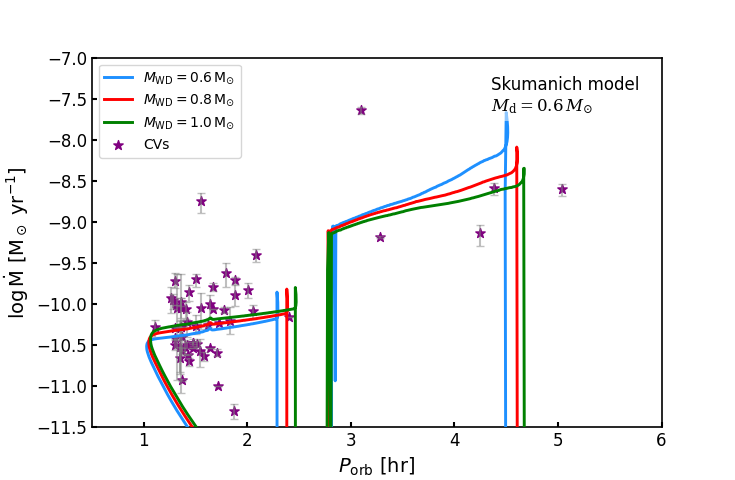}
  \includegraphics[width=0.48\textwidth]{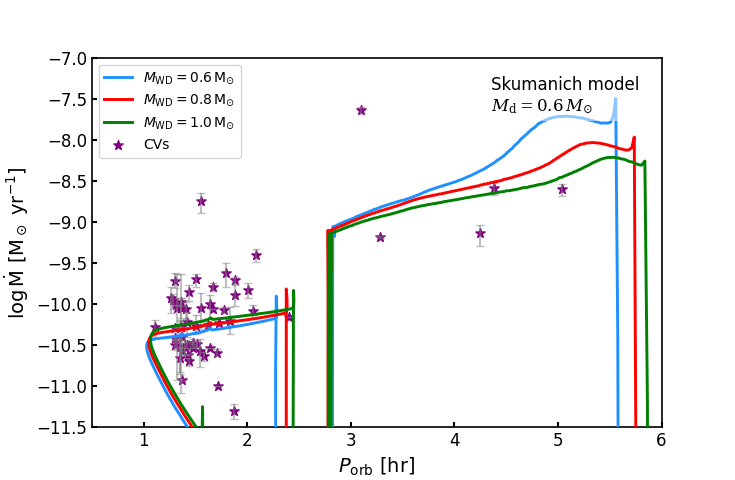}
  
  \vspace{0.3cm}
  
  \includegraphics[width=0.48\textwidth]{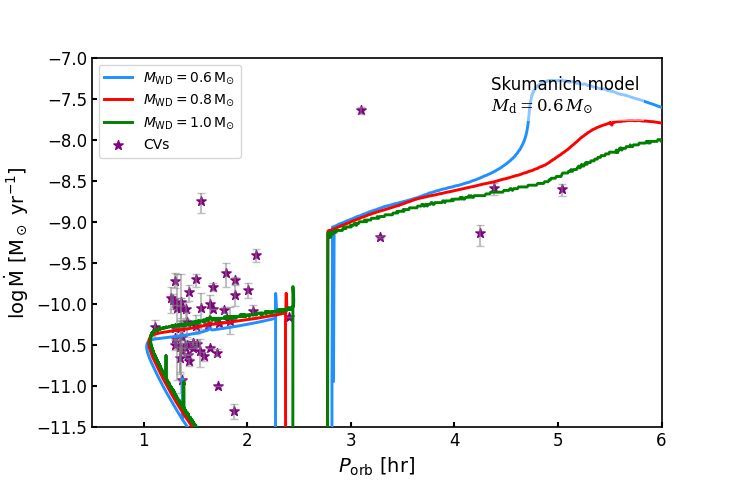}
  \includegraphics[width=0.48\textwidth]{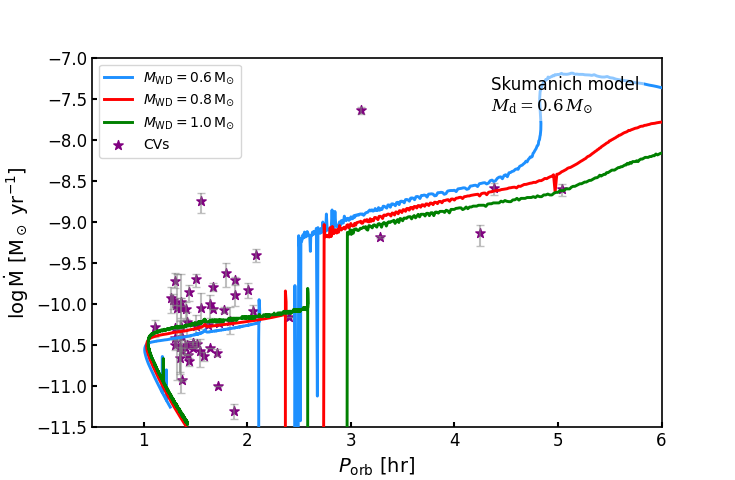}
  
  \caption{Evolutional tracks of the Skumanich model: mass-transfer rate as a function of orbital period for different donor masses—0.6 $M_\odot$ (top left), 0.8 $M_\odot$ (top right), 1.0 $M_\odot$ (bottom left), and 1.2 $M_\odot$ (bottom right). Each panel shows the evolutionary tracks for three different WD masses: 0.6 $M_\odot$ (blue), 0.8 $M_\odot$ (red), and 1.0 $M_\odot$ (green). The purple pentagrams denote observational data obtained from \citet{Pala2020}, \citet{Pala2022}, and \citet{Sarkar2024}, as well as the additional sources listed in Table~A.1. All systems are initialized at an orbital period of 0.4~day.}
  \label{fig:sku_model}
\end{figure*}

\begin{figure*}[ht!]
  \centering
  \includegraphics[width=0.48\textwidth]{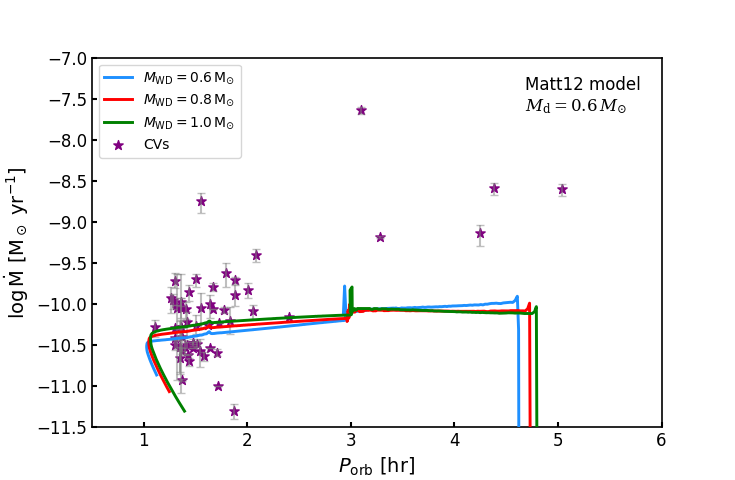}
  \includegraphics[width=0.48\textwidth]{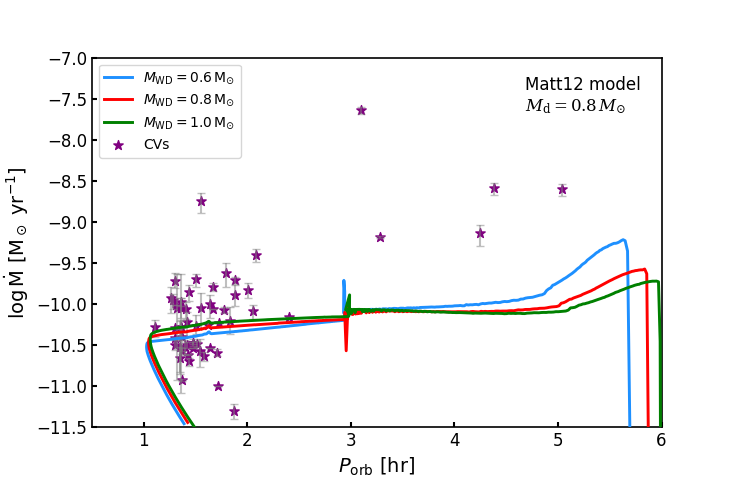}
  \vspace{0.3cm}
  \includegraphics[width=0.48\textwidth]{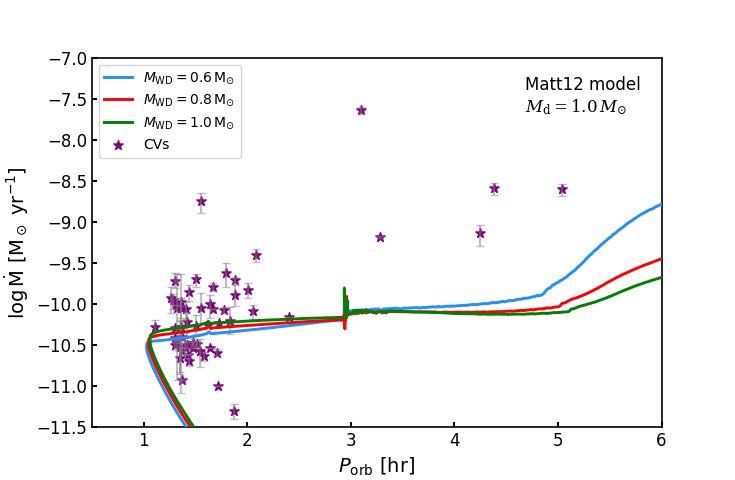}
  \includegraphics[width=0.48\textwidth]{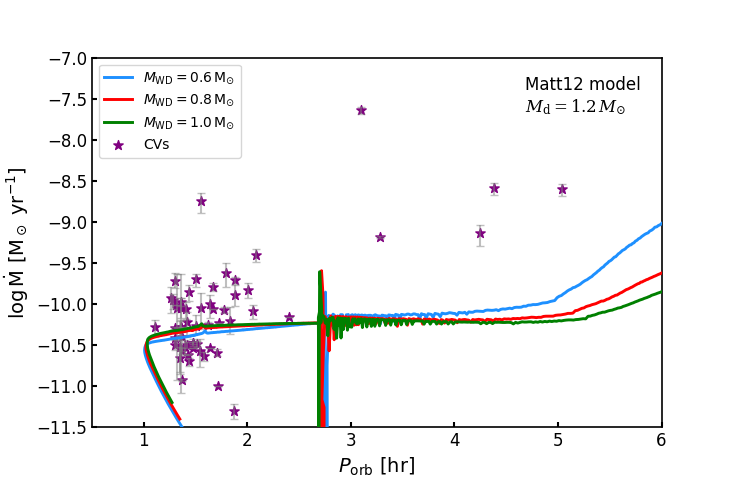}  
  \caption{Same as Fig.~\ref{fig:sku_model} but for the Matt12 model.}
  \label{fig:matt_model}
\end{figure*}

\subsubsection{Matt model (Matt12 model)}
In order to more accurately describe the AML mechanisms of solar-type stars under different magnetic field strengths and wind parameters, \citet{Matt2012} proposed a new picture of AML driven by the interaction between stellar winds and magnetic fields, based on magnetohydrodynamic simulations of stellar winds. The stellar wind does not flow out uniformly in all directions but is confined near magnetic field lines and carries away angular momentum along these lines. The stronger the magnetic field, the longer the angular momentum ``lever arm,'' and the more efficient the loss:
\begin{equation}
\dot{J}_{\rm MB} = -\frac{K_1^2}{(2G)^{m}} \overline{B}_s^{4m}\dot{M}_{1,w}^{1-2m}
\left( \frac{R_1^{5m+2}}{M_1^m} \right)
\left( \frac{\Omega}{(K_2^2 + 0.5u^2)^m} \right)
,\end{equation}
where \(u = \sqrt{2GM/R}\) is the surface escape velocity of the star. The calibration constants are \(K_1 = 6.7\), \(K_2 = 0.506\), and the parameter \(m = 0.17\), and is related to the magnetic field geometry. The average magnetic field strength is given by \(\overline{B}_s = fB_s\), where \(f\) is the filling factor of the magnetized region on the stellar surface, and \(B_s = B_\odot = 1 \, \text{G}\) is the surface magnetic field strength of the Sun, all based on the work of \citet{Gallet2013}.

The magnetic field filling factor, \(f\), which expresses the magnetized fraction of the stellar surface \citep{Saar1996, Amard2016}, depends on the normalized Rossby number, which is expressed as
\begin{equation}
f = \frac{0.4}{\left[1 + \left(\frac{x}{0.16}\right)^{2.3}\right]^{1.22}}
, \end{equation}
This expression introduces MB saturation at high rotation rates, where the magnetic field strength does not increase indefinitely and the filling factor approaches saturation. This feature is supported by observations of rapidly rotating stars in young clusters \citep{Wright2011}. The normalized Rossby number is defined as
\begin{equation}
x = \left( \frac{\Omega_\odot}{\Omega} \right) 
\left( \frac{\tau_{\rm conv}}{\tau_{\odot,\rm conv}} \right)
, \end{equation}
where \(\Omega_\odot = 3 \times 10^{-6} \, \text{s}^{-1}\) is the solar angular velocity, and \(\tau_{\odot,\rm conv} = 2.8 \times 10^6 \, \text{s}\) is the solar convective turnover time. A smaller Rossby number indicates higher dynamo efficiency, i.e., a stronger magnetic field.

The stellar wind loss rate (\(\dot{M}_{1,w}\)), which describes the steady-state wind loss driven by the star itself, was calculated using the stellar wind mass-loss formula of \citet{Reimers1975}:
\begin{equation}
\dot{M}_{1,w} = 4 \times 10^{-13} \eta \left( \frac{L_1}{L_\odot} \right) \left( \frac{R_1}{R_\odot} \right) \left( \frac{M_1}{M_\odot} \right)^{-1} \, M_\odot \, \text{yr}^{-1}
,\end{equation}
where \(\eta\) is a scaling or efficiency parameter for the wind loss, we adopt \(\eta = 1\). \(L_1\) is the luminosity of the donor star.

\subsubsection{Reiners \& Mohanty model (RM12 model)}
Many rapidly rotating low-mass stars observed in young clusters maintain high rotation rates for extended periods, suggesting that the MB mechanism is significantly weakened or enters a saturated regime in these stars. \citet{Reiners2012} constructed a ``magnetic saturation, radius-dominated'' AML model (hereafter the RM12 model), attempting to introduce a strong radius dependence (\(R^{16/3}\)), emphasizing the structural dependence of the MB mechanism in low-mass stars and the nonlinear saturation behavior of rotational braking under strong magnetic field conditions. The mathematical expression for AML in this model is

\begin{equation}
\frac{dJ}{dt} = 
\begin{cases}
- \mathcal{C} \left[ \Omega \left( \frac{R^{16}}{M^2} \right)^{1/3} \right], & \text{if } \Omega \geq \Omega_{\text{sat}} \\
- \mathcal{C} \left[ \left( \frac{\Omega}{\Omega_{\text{sat}}} \right)^4 \Omega \left( \frac{R^{16}}{M^2} \right)^{1/3} \right], & \text{if } \Omega < \Omega_{\text{sat}}
\end{cases}
,\end{equation}
Here \(\Omega\) denotes the stellar angular velocity, and \(\Omega_{\text{sat}}\) is the critical threshold for saturation. Following the formulation of \citet{Reiners2012} and \citet{Amard2016}, the proportionality constant \(\mathcal{C}\) can be explicitly expressed as
\begin{equation}
\mathcal{C} = \frac{2}{3} \left( \frac{B_{\mathrm{crit}}^8}{G^2 K_V^4 \dot{M}^2} \right)^{1/3}
,\end{equation} Here, $B_{\mathrm{crit}}$ is the critical magnetic field strength marking the onset of dynamo saturation. The factor $K_V$ is some dimensionless constant scaling factor.

It is worth noting that the factor \(\mathcal{C}\) includes several parameters with considerable uncertainties. Observations of M dwarfs with saturated magnetic fields and very young T Tauri stars indicate that \(B_{\text{crit}}\) is roughly at the level of a few thousand gauss; however, in this field, data for field stars are both scarce and weakly constrained \citep{Saar1996}. From a theoretical perspective, to ensure the validity of the escape velocity assumption, the wind parameter (\(K\)) should be close to 1, although this has yet to be confirmed observationally. Meanwhile, the wind mass loss rate (\(\dot{M}\)) is typically set to the present-day solar value of approximately \(10^{-14}\,M_{\odot}\,\mathrm{yr}^{-1}\); however, for stars of different masses and ages, this parameter remains highly uncertain. Recent simulation studies suggest that in very low-mass stars, \(\dot{M}\) could be several orders of magnitude higher than this value \citep{Vidotto2011}. 

Because each of these quantities carries significant uncertainty, \(\mathcal{C}\) itself may vary by several orders of magnitude. To evaluate the impact of this uncertainty, we examined how variations in \(\mathcal{C}\) affect the predicted AML rate. Increasing \(\mathcal{C}\) by an order of magnitude enhances the MB torque and produces spin-down timescales more consistent with the observed period gap in close binaries, whereas smaller values lead to weaker braking and slower angular momentum evolution. We adopted the calibration proposed by \citet{Amard2016}, setting \(\mathcal{C} = 10^{39}\ \mathrm{g^{5/3}\,cm^{-10/3}\,s}\) and the saturation threshold \(\Omega_{\text{sat}} = 3\,\Omega_\odot\).

\begin{figure*}[ht!]
  \centering
  \includegraphics[width=0.48\textwidth]{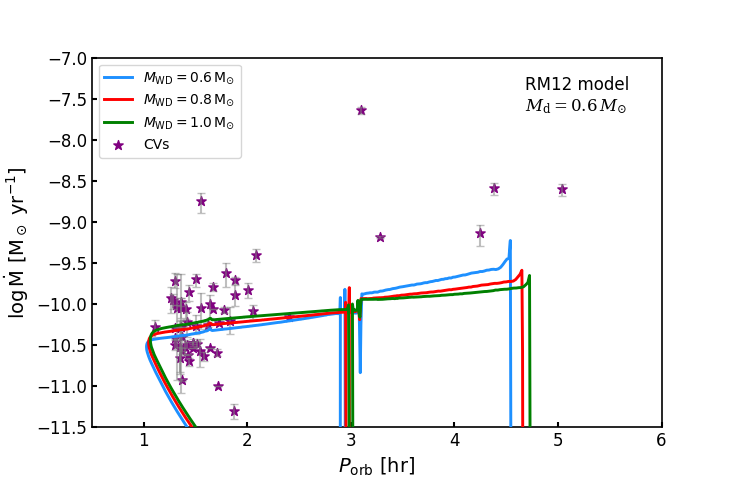}
  \includegraphics[width=0.48\textwidth]{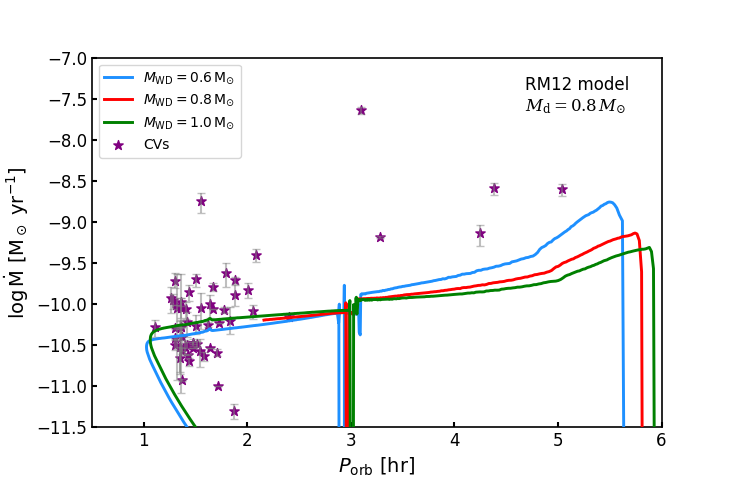}
  
  \vspace{0.3cm}
  
  \includegraphics[width=0.48\textwidth]{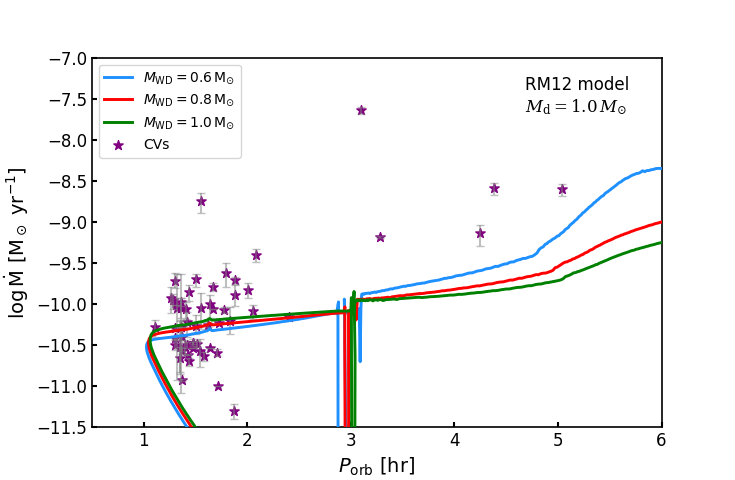}
  \includegraphics[width=0.48\textwidth]{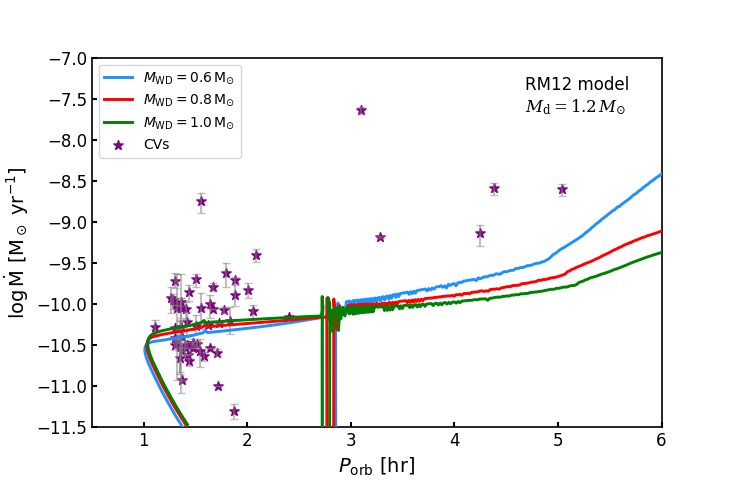}
  
  \caption{Same as Fig.~\ref{fig:sku_model} but for the RM12 model.}
  \label{fig:RM12_model}
\end{figure*}

\subsubsection{Intermediate model}
The intermediate model was initially proposed by \citet{Rappaport1983} to resolve the discrepancy where the traditional Skumanich MB law underestimates the observed mass-transfer rates by orders of magnitude in low- and intermediate-mass LMXBs. This model incorporates several scaling enhancement factors, including convective turnover time, rotation rate, and stellar wind mass loss. The AML rate is given by
\begin{equation}
\dot{J}_\mathrm{MB} = \dot{J}_\mathrm{MB,Sk} \left( \frac{\Omega}{\Omega_\odot} \right)^\beta \left( \frac{\tau_\mathrm{conv}}{\tau_{\odot,\mathrm{conv}}} \right)^\xi \left( \frac{\dot{M}_\mathrm{W}}{\dot{M}_{\odot,\mathrm{W}}} \right)^\alpha,
\end{equation}
where the exponents are $\beta=0$, $\xi=2$, and $\alpha=1$. This formulation follows the approach proposed by \citet{Van2019}, providing a more flexible scheme that accounts for the expected enhanced AML in stars with strong convection and increased mass loss, such as subgiants or red giants in close binaries.

\subsubsection{Convection-boosted model (boosted model)}
In this model, we consider only the convective enhancement to MB, ignoring the direct effect of wind mass-loss. This prescription proposed by \citet{Van2019} has been shown to better reproduce persistent LMXB systems compared to the default model, but its application to the evolution of CVs has not yet been investigated. This approach is appropriate for stars with significant convective envelopes but weak winds, such as low-mass main-sequence stars. It uses the same general form as the intermediate model, but with modified exponents: $\beta = 0$, $\xi = 2$, and $\alpha = 0$. This formulation emphasizes the role of the convective turnover time in strengthening the magnetic field via the dynamo mechanism, as suggested by \citet{ivanova2006} and supported by observational trends in \citet{auriere2015}.

\begin{figure*}[ht!]
  \centering
  \includegraphics[width=0.48\textwidth]{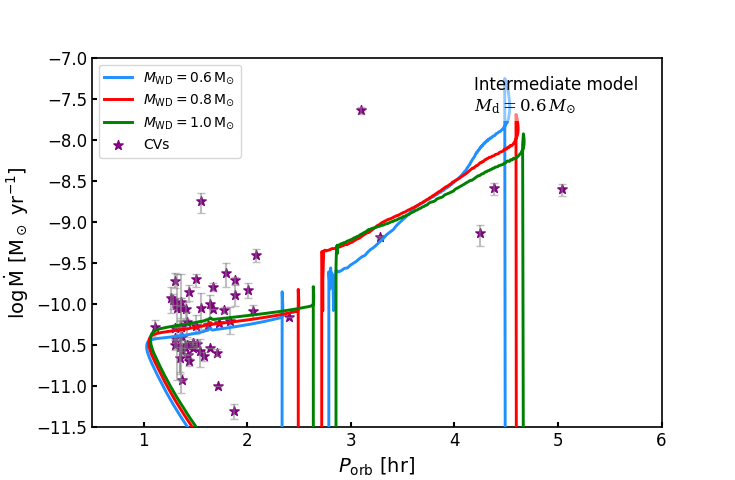}
  \includegraphics[width=0.48\textwidth]{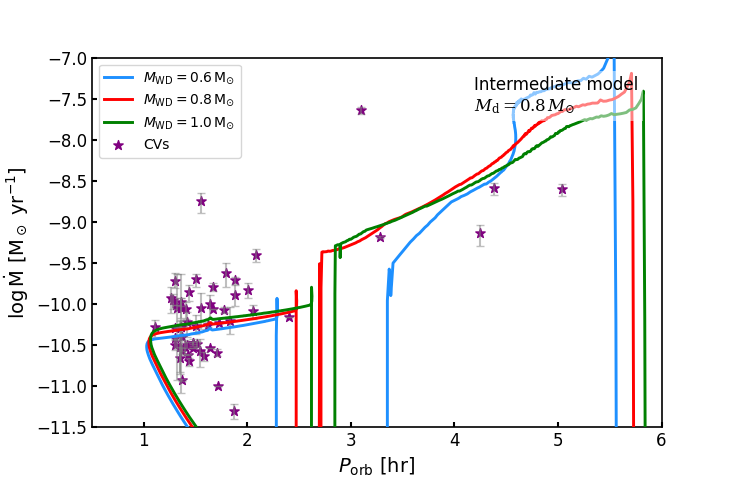}
  \vspace{0.3cm}
  \includegraphics[width=0.48\textwidth]{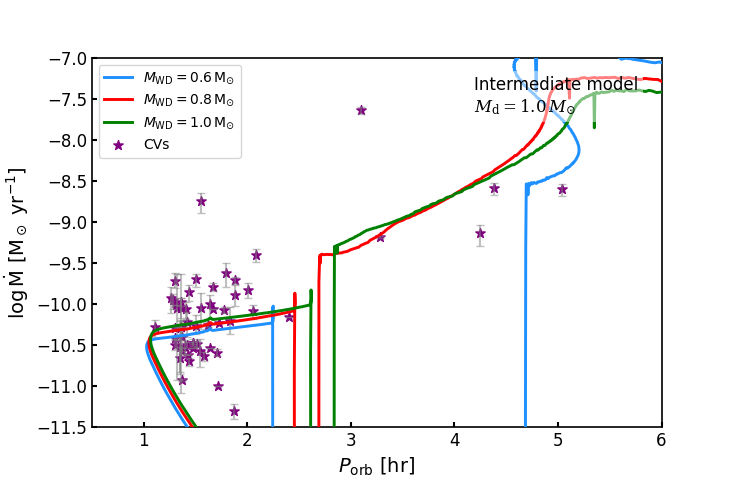}
  \includegraphics[width=0.48\textwidth]{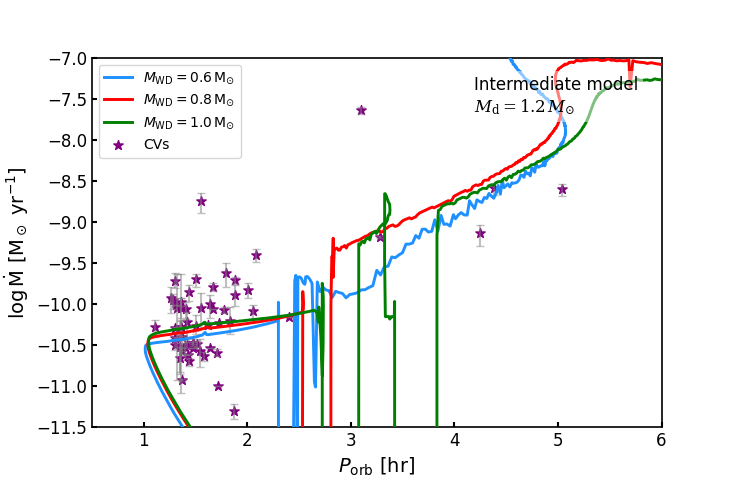}
  \caption{Same as Fig.~\ref{fig:sku_model} but for the intermediate model.}
  \label{fig:inter_model}
\end{figure*}

\section{Results}
In this section we present the results of our binary evolution calculations. We focus on the evolutionary tracks of CVs, the comparison between our model predictions and observational data, and the impact of different magnetic field assumptions on the evolution of the systems.

\begin{figure*}[ht!]
  \centering
  \includegraphics[width=0.48\textwidth]{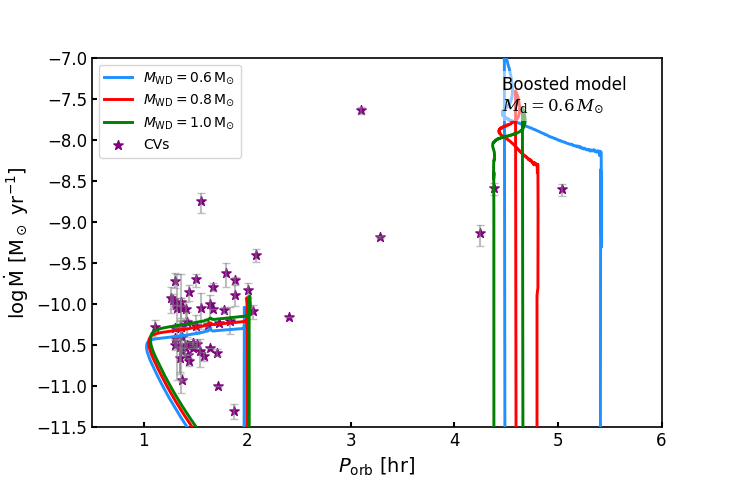}
  \includegraphics[width=0.48\textwidth]{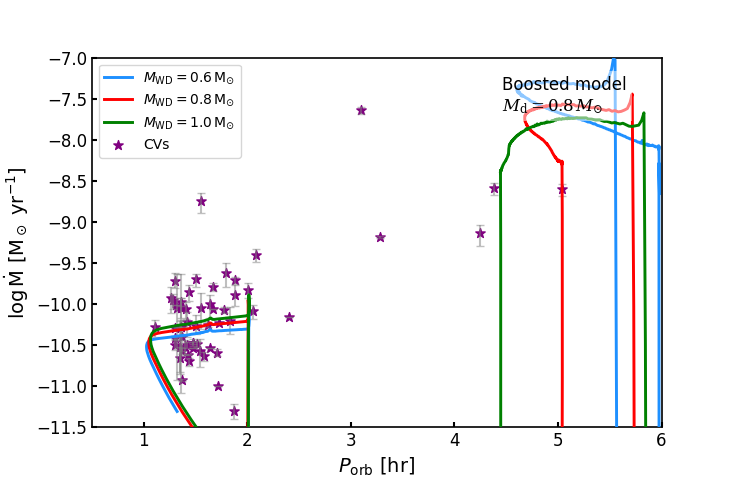}
  
  \vspace{0.3cm}
  
  \includegraphics[width=0.48\textwidth]{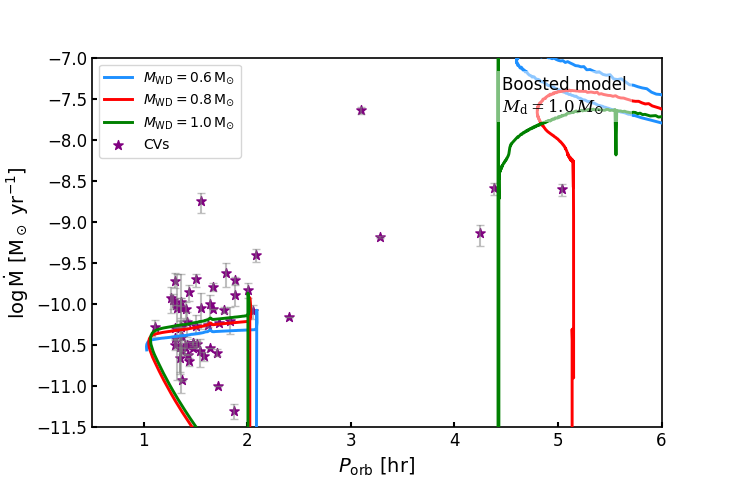}
  \includegraphics[width=0.48\textwidth]{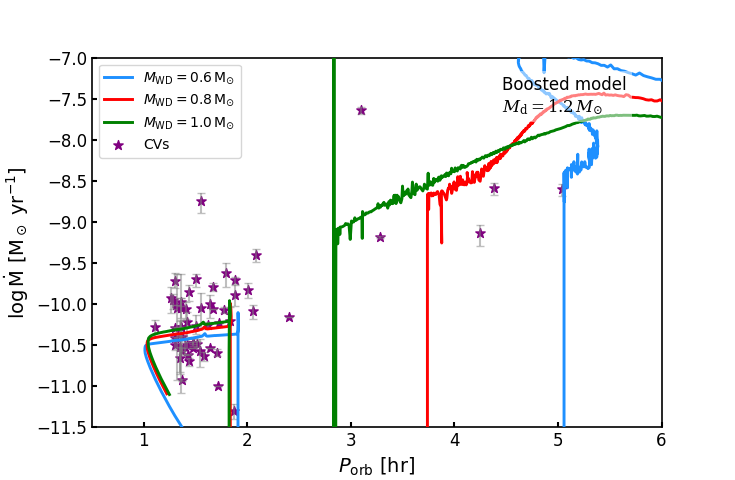}
  
  \caption{Same as Fig.~\ref{fig:sku_model} but for the boosted model.}
  \label{fig:Boosted_model}
\end{figure*} 

\begin{figure*}[ht!]
  \centering

  \begin{minipage}[t]{0.48\textwidth}
    \centering
    \includegraphics[width=\linewidth]{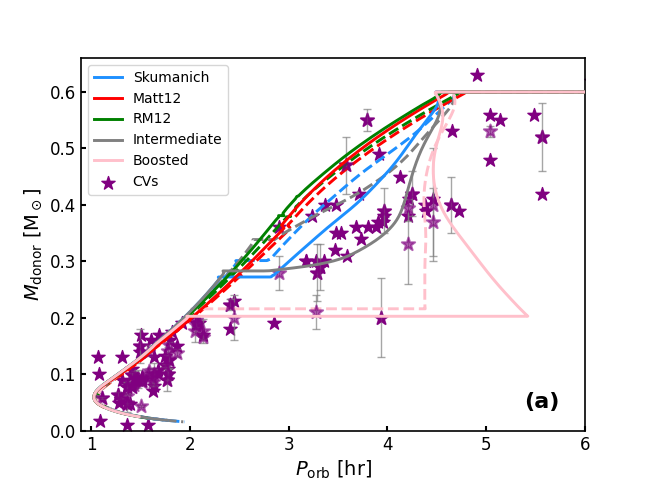}
    \label{fig:g1_mass}
  \end{minipage}
  \hfill
  \begin{minipage}[t]{0.48\textwidth}
    \centering
    \includegraphics[width=\linewidth]{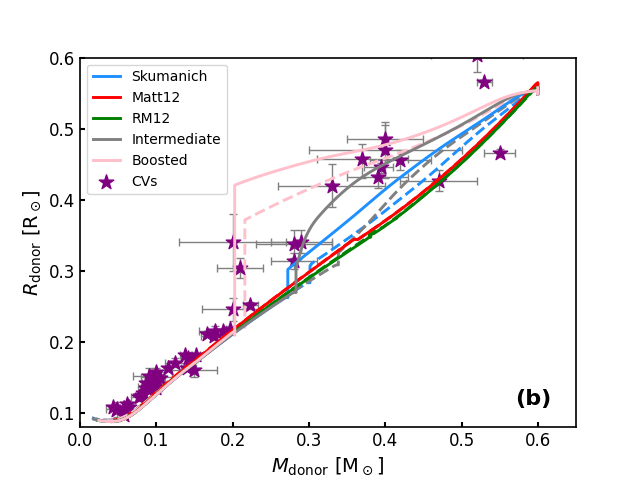}
    \label{fig:g1_mass_radius}
  \end{minipage}

  \vspace{0.2cm}
  \begin{minipage}[t]{0.48\textwidth}
    \centering
    \includegraphics[width=\linewidth]{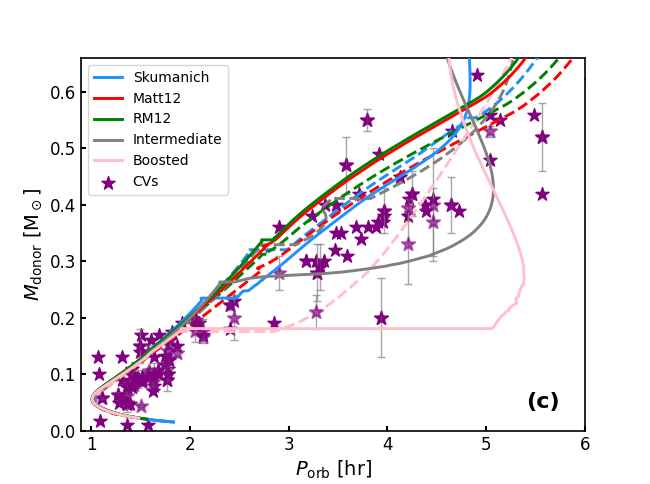}
    \label{fig:g2_mass}
  \end{minipage}
  \hfill
  \begin{minipage}[t]{0.48\textwidth}
    \centering
    \includegraphics[width=\linewidth]{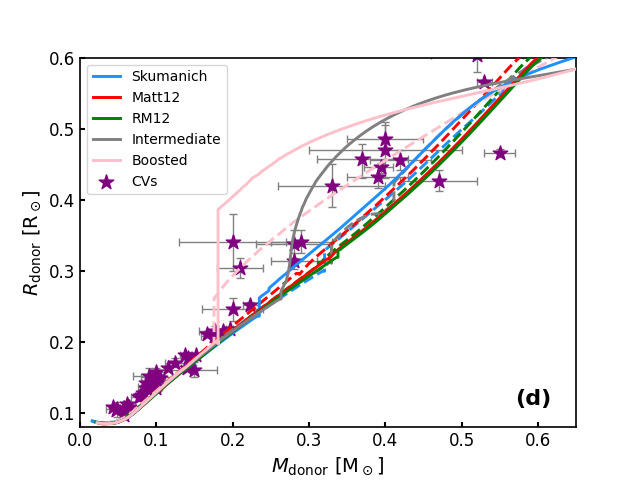}
    \label{fig:g2_mass_radius}
  \end{minipage}

  \caption{Evolutionary diagrams comparing orbital period ($P_{\rm orb}$) and donor star properties for five MB prescriptions. Left panels: Relationship between $P_{\rm orb}$ and donor mass ($M_{\rm donor}$). Right panels: Corresponding donor radius ($R_{\rm donor}$) versus mass evolution. In panels (a) and (b), we show $M_\mathrm{donor,i}=0.6\,M_\odot$ + $M_\mathrm{WD,i}=0.6\,M_\odot$ (solid lines) and  $M_\mathrm{donor,i}=0.6\,M_\odot$ + $M_\mathrm{WD,i}=1.0\,M_\odot$ (dashed). (b) In panels (c) and (d) we show $M_\mathrm{donor,i}=1.2\,M_\odot$ + $M_\mathrm{WD,i}=0.6\,M_\odot$ (solid) and $M_\mathrm{donor,i}=1.2\,M_\odot$ + $M_\mathrm{WD,i}=1.0\,M_\odot$ (dashed). All models share initial orbital period $P_\mathrm{orb,i}=0.4$\,d. Observed systems from \citet{McAllister2019} are marked with purple stars.}
  \label{fig:set1}
\end{figure*}

\begin{figure}[htbp]
  \centering
  \hspace*{-0.6cm}
  \includegraphics[scale=0.53]{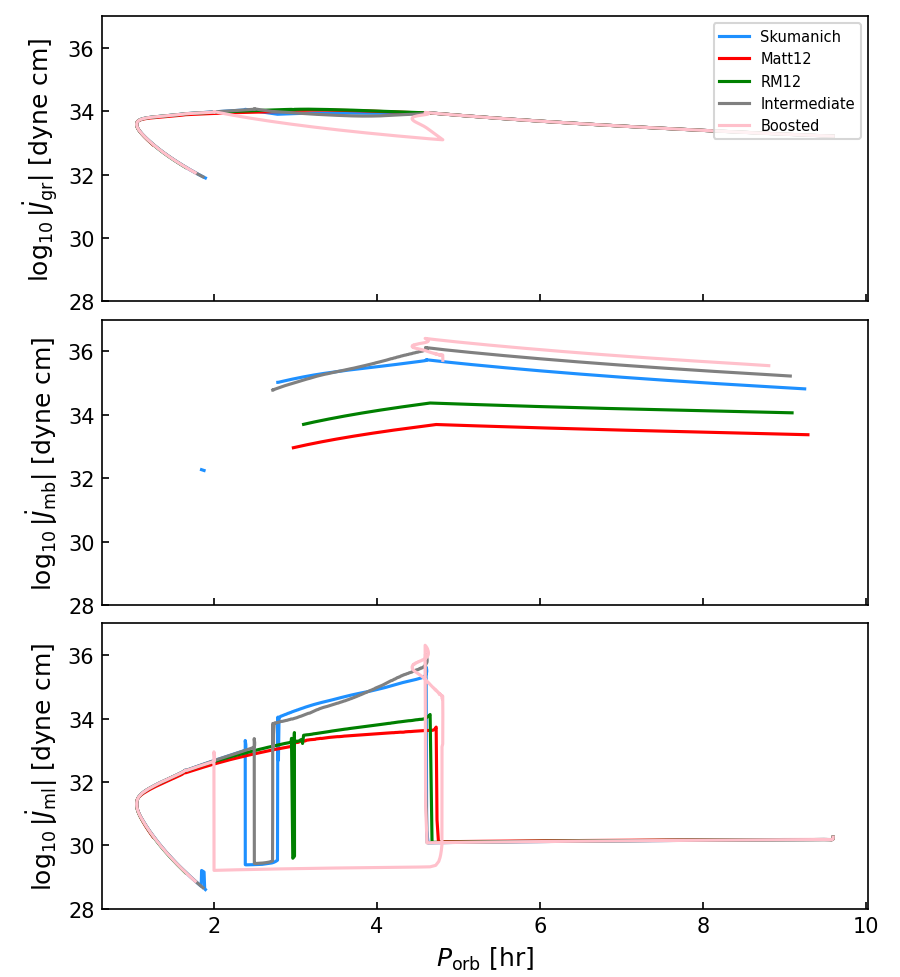}
  \caption{Relationship between the AML rate caused by MB ($\dot{J}_{\text{mb}}$), gravitational wave radiation ($\dot{J}_{\text{gr}}$), and mass loss ($\dot{J}_{\text{ml}}$) and the orbital period in the evolution process of the five models. The models are shown in different colors: Skumanich (blue), Matt12 (red), RM12 (green), intermediate (gray), and boosted (pink).}
  \label{fig:AML}
\end{figure} 

\begin{figure}[htbp]
  \centering
  \hspace*{-0.3cm} 
  \includegraphics[width=1\linewidth]{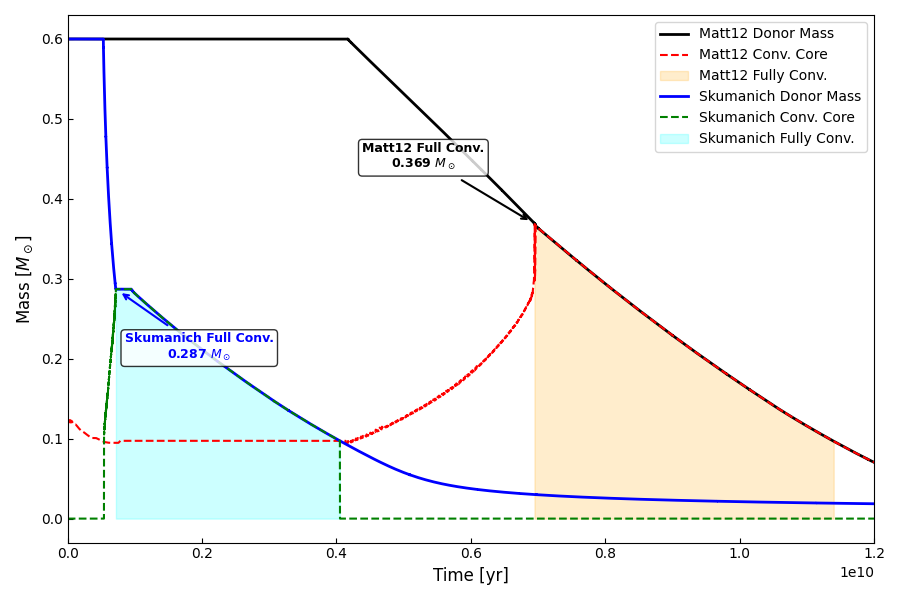}
  \caption{Evolution of donor mass and convective properties in Matt12 and Skumanich models. Solid lines show the donor mass evolution, while dashed lines represent the convective core mass. Orange and cyan shaded regions indicate periods of fully convective envelopes for the Matt12 and Skumanich models, respectively. Black and blue arrows mark the onset of fully convective states, with corresponding mass values labeled. }
  \label{fig:Convection}
\end{figure} 

\begin{figure}[htbp]
  \centering
  \includegraphics[width=0.48\textwidth]{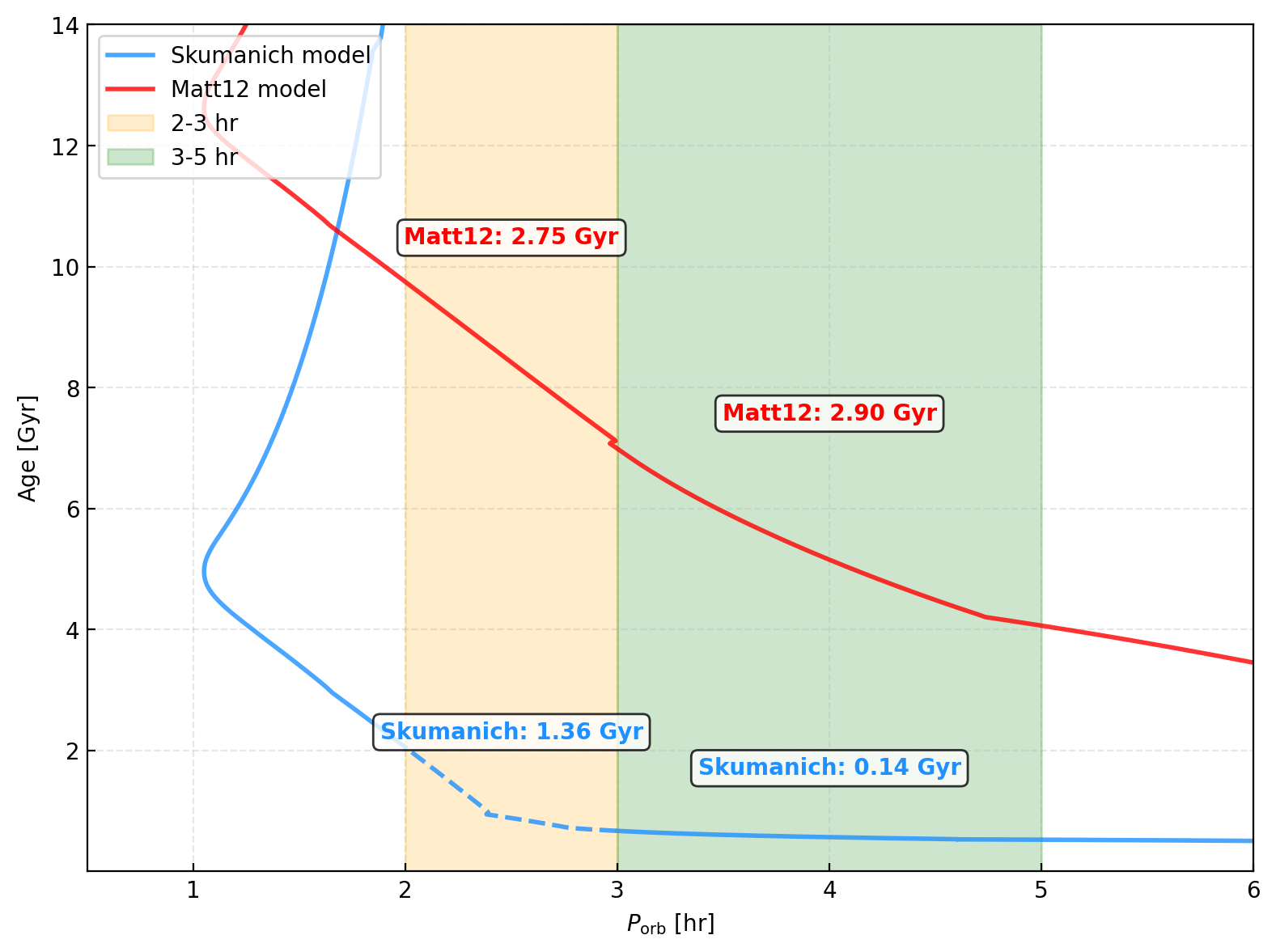}
  \caption{Orbital period versus timescale relationships for the Skumanich (blue) and Matt12 (red) models. The solid lines represent their respective evolutionary tracks. The timescales experienced during evolution in the orbital period ranges of 2--3 hours and 3--5 hours are labeled for both models.}
  \label{fig:Timescale}
\end{figure}

\begin{figure}[htbp]
  \centering
  \includegraphics[width=0.48\textwidth]{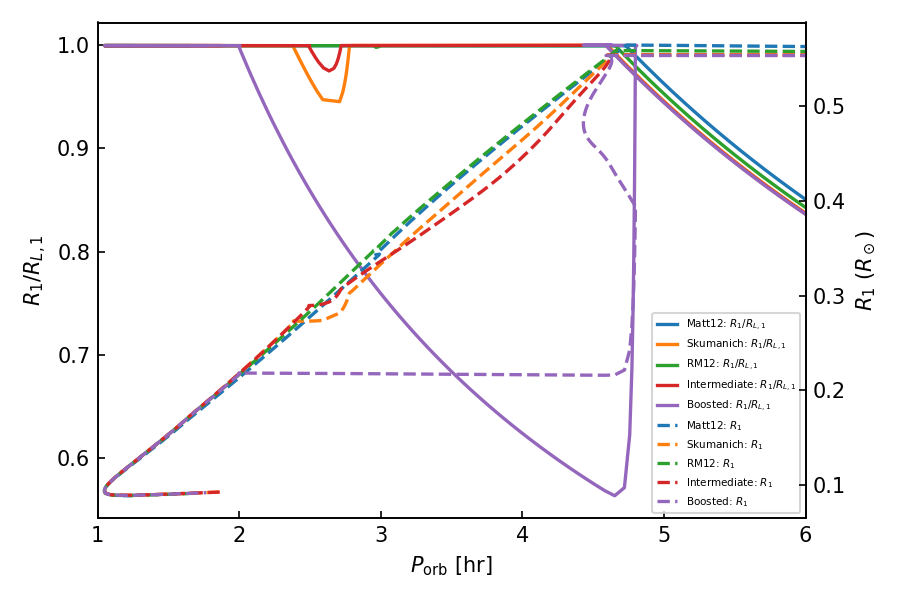}
  \caption{Relations between orbital period, donor radius ($R_1$, dashed lines), and radius ratio ($R_1/R_{L,1}$, solid lines), with models distinguished by color: Matt12 (blue), Skumanich (orange), RM12 (green), intermediate (red), and boosted (purple).}
  \label{fig:RLOF}
\end{figure}

\subsection{Evolutionary tracks}
Figure~\ref{fig:sku_model} illustrates the evolutionary tracks of the mass-transfer rate as a function of orbital period for binary systems with varying initial donor masses ($0.6$, $0.8$, $1.0$, and $1.2\,M_{\odot}$) and different WD masses ($0.6$, $0.8$, and $1.0\,M_{\odot}$) by MB of Skumanich model. Each curve corresponds to a distinct WD mass, depicting the theoretical prediction of the mass-transfer rate of the donor along the orbital evolution. All systems are initialized with an identical orbital period of 0.4 days to ensure direct comparability.

For the models with lower-mass donors (e.g., \( M_{\mathrm{d}} = 0.6 \) and \( 0.8\,M_{\odot} \)) by MB of Skumanich model, WD mass has only a minor impact on the evolutionary outcome near the upper edge of the period gap, with the width of the period gap remaining nearly consistent. Regardless of WD mass, these systems follow similar evolutionary tracks and tend to enter the period gap at comparable orbital periods. This behavior is primarily attributed to the fully convective or deep convective structure of low-mass donors, which renders their stellar radius relatively insensitive to mass loss, thereby diminishing the influence of WD mass on orbital evolution \citep{Knigge2011, Rappaport1983}.

In contrast, when the initial donor mass increases (e.g., \( M_{\mathrm{d}} = 1.2\,M_{\odot}\)), the WD mass plays a far more significant role in shaping the evolutionary path. This heightened sensitivity can be understood through mass-transfer stability analysis: for a given systemic AML rate, a larger initial mass ratio \( q = M_{\mathrm{d}} / M_{\mathrm{WD}} \) leads to a higher mass-transfer rate \citep{Ge2024, Li2025}, and the width of the period gap is affected: the models with more massive WDs reach the period gap earlier and exhibit steeper declines in the mass-transfer rate. This sensitivity highlights that, in the models with more massive donors, the WD mass has a certain influence on the timescale and orbital period at which the system enters the gap, but its overall effect is relatively limited, because the WD mass affects the timescale of orbital evolution within the period gap, with the influence proportional to \( M_{\mathrm{WD}}^{2/3} \) \citep{Paczynski1981}. Once the system has entered the period gap and MB has ceased, the subsequent orbital evolution is governed primarily by gravitational radiation (GR). In this detached phase, a more massive WD produces slightly stronger gravitational wave emission, which leads to faster AML and causes the binary to shrink more rapidly toward the lower edge of the gap, around $P_{\mathrm{orb}} \sim 2$ hours \citep{Landau1975}. Nevertheless, this moderate dependence during the gap does not imply that the WD mass plays a dominant role in shaping the global properties of the CV population. Although the WD mass exerts some influence on orbital evolution, the overall characteristics of CV systems (such as the orbital period distribution) are influenced by multiple factors, including the binary evolution process and the initial conditions of the system, and are not solely determined by the WD mass. Therefore, while the WD mass affects the orbital evolution of individual systems, a clearer understanding of its role in the CV population requires further investigation through population synthesis studies.

Lastly, the Skumanich model curves in Fig.~\ref{fig:sku_model} clearly reproduce a pronounced period gap and show partial agreement with observational data. In the short-period regime ($P_{\mathrm{orb}} < 2$ hours), a significant number of observed systems appear to be period bouncers. The Skumanich model successfully explains part of this population. However, the model fails to fully account for the broad distribution of observed systems in this regime. In the longer-period regime ($P_{\mathrm{orb}} > 3$ hours), observational samples are sparser, but the predicted mass-transfer rates show reasonable consistency with the data. We believe that when comparing observed and calculated mass-transfer rates, it should be noted that the MESA model calculates the secular mean mass-transfer rate, while the observed mass-transfer rates typically reflect instantaneous measurements. Even when based on accurate fluxes and distances, observed values may still be influenced by periodic variations or other physical phenomena \citep{Siwak2018}. These discrepancies suggest that, despite the general predictive power of the model, additional physical processes—such as enhanced AML due to magnetic activity \citep{Matt2012, Gossage2021} or consequential AML following nova eruptions (e.g., \citealt{Tang2024})—might influence the long-term evolution of CVs and should be incorporated in future modeling.

Figure~\ref{fig:matt_model} shows the evolutionary tracks of the Matt12 model. Obviously, this model fails to reproduce the characteristic period gap of CVs. In the Matt12 model, even after the donor becomes fully convective and MB shuts off, a substantial AML rate remains. Consequently, the orbit kepng shrinking and the donor stays in permanent Roche-lobe contact. Therefore, the Matt12 model may be more appropriate for magnetic CVs, specifically polar and IP systems (a detailed discussion appears in Sect. 4).

As shown in Fig.~\ref{fig:RM12_model}, the RM12 model predicts a brief detached phase near an orbital period of about 3 hours, and no significant period gap. The discrepancy with the observational results may arise from the choice of parameters in the model. Specifically, the normalization constant \( C \), stellar mass \( M \), and critical magnetic field strength \( B_{\mathrm{crit}} \) are calibrated based on observational data or the spin evolution fits of isolated main-sequence stars (such as the Sun or M dwarfs; \citealt{Reiners2012, Matt2015}). However, these assumptions may not be applicable to CV systems. The AML in binary systems could differ significantly from single-star models due to changes in the magnetic field structure and AML mechanisms caused by tidal synchronization, mass transfer, and other complex interactions \citep{Knigge2011,Garraffo2018,Kovari2024}. Therefore, the RM12 model's failure to reproduce the period gap may reflect the limitations of applying single-star magnetic presets to mass transfer binary systems.

Figure~\ref{fig:inter_model} illustrates the intermediate model, which exhibits a significantly stronger dependence of the orbital period gap width on the WD mass compared to the Skumanich model. Additionally, the intermediate model generally predicts higher systemic mass-transfer rates than the Skumanich model. This discrepancy may arise from the incorporation of the convective turnover timescale and an enhanced stellar wind mass-transfer rate in the AML prescription of the intermediate model. These factors are particularly impactful for stars with deep convective envelopes and higher wind mass-transfer rates.

The boosted model shown in Fig.~\ref{fig:Boosted_model} exhibits the widest period gap among all the models considered. For a system with an initial donor mass of \(1.2\,M_\odot\), the initial mass of the WD plays a crucial role in determining the width of the period gap. For instance, when the initial WD mass is \(0.6\,M_\odot\), the resulting period gap width is approximately three times that of a system with a \(1.0\,M_\odot\) WD. This significant difference primarily arises from variations in the mass ratio, defined as \(q = M_{\mathrm{donor}} / M_{\mathrm{WD}}\). A lower initial WD mass corresponds to a higher mass ratio, which causes the donor star to develop a deeper convective envelope as it loses mass. A deeper convective envelope enhances the efficiency of the MB via the magnetic hysteresis mechanism—whose additional influence depends mainly on the convective turnover time. Once the system separates, the rate of orbital shrinkage is primarily determined by GR. In fact, the wider period gap in the boosted model is caused by two factors: on one hand, the donor is more inflated when MB is disrupted, resulting in a longer orbital period at detachment; on the other hand, the donor has a smaller equilibrium radius, which leads to a shorter orbital period when mass transfer resumes \citep{Zorotovic2016}. The combined effect of these two factors results in a significantly wider period gap.

Figure~\ref{fig:set1} presents the evolutionary tracks of CVs in the orbital period-companion mass and companion mass-radius planes, respectively. Compared with observational data, the three models - Skumanich, intermediate, and boosted - can generally cover the observed distribution of companion masses. Among them, the boosted model performs best in explaining systems with initially low-mass companions in the short-period range of 2--5 hours, while the Skumanich and intermediate models are more suitable for describing systems with initially more massive companions in the same period range. For systems within the period gap, both the intermediate and boosted models provide better agreement with observations, particularly for systems with initial companion masses of \(0.6\,M_\odot\), where the consistency is most pronounced. In contrast, the Matt12 and RM12 models (after excluding period bouncer candidates) often show larger discrepancies with observed systems. We suggest that the intermediate model, which does not rely on a single mechanism (such as the empirical spin-down law in the Skumanich model or the enhanced convection in the boosted model), is more appropriate for describing the long-term evolution of such binary systems. By comprehensively considering the mass-transfer rate, period gap characteristics, and changes in companion mass and radius, this model can cover and explain a wider range of observational data.

\subsection{The influence of magnetic fields on period gaps}
In the Matt12 model, the role of surface magnetic fields in donor stars influences the evolution of orbital period and mass-transfer rate in CV systems. As $B_s$ increases, stronger magnetic fields enhance the MB effect, increasing AML and accelerating the mass-transfer rate. Although the regulatory effect of surface magnetic fields on the evolution of the system is limited, their contribution to the enhancement of MB, and thus the resulting orbital contraction and increase in mass-transfer rate, remains crucial.

To investigate why some models fail to reproduce the period gap, we compared the AML rates due to MB, gravitational radiation, and mass loss during the evolution of several models. As shown in Fig.~\ref{fig:AML}, for $\dot{J}_{\mathrm{MB}}$, different models exhibit magnitude differences. For instance, the Skumanich model yields values roughly two orders of magnitude higher than the Matt12 model, indicating higher MB efficiency. Models with stronger MB can remove angular momentum more efficiently at longer orbital periods, leading to faster orbital contraction and quicker evolution of the donor star toward shorter periods. This discrepancy may arise because traditional models like Skumanich model do not incorporate mechanisms related to magnetic field self-regulation and structural adjustment, whereas the Matt12 and RM12 models introduce magnetic-field-dependent parameters and dynamic Alfv\'en radius adjustment. In high-rotation environments, wind acceleration driven by magneto-centrifugal effects reduces the Alfv\'en radius, thereby suppressing excessive AML.

Meanwhile, near an orbital period of 3 hours, all five models exhibit a sudden drop in \( \dot{J}_{\rm MB} \) to nearly zero. When the orbital period shrinks to about 3 hours, the low-mass donor star reaches a fully convective state due to continued radius contraction and internal structural evolution—that is, the entire star becomes convective with no radiative core~\citep{Rappaport1983, spruit1983}. Once the donor becomes fully convective, the magnetic field structure required for efficient MB is thought to collapse or weaken sharply, resulting in the shutdown of the MB mechanism~\citep{Hussain2011, schreiber2010}. This is the common physical reason for the sudden drop in \( \dot{J}_{\rm MB} \) in both models near the 3-hour period. 

We compare two representative models as shown in Fig.~\ref{fig:Convection}, where the time and mass at which the companion star reaches full convection differ significantly between models: approximately 0.29 \(M_\odot\) in the Skumanich model while 0.37 \(M_\odot\) in the Matt12 model. We further compared the evolutionary timescales of both models in the 2--3 hour and 3--5 hour orbital period ranges, as illustrated in Fig.~\ref{fig:Timescale}. In the 2--3 hour period gap region, the evolutionary timescales are 1.36 Gyr for Skumanich and 2.75 Gyr for Matt12. For the 3--5 hour range, the Skumanich model shows a much shorter timescale of only 0.14 Gyr compared to 2.90 Gyr for Matt12. We attribute this mass difference to the stronger MB in the Skumanich model. The higher mass-transfer rate not only implies faster stripping of surface material from the donor, but also drives the donor out of thermal equilibrium, leading to a more rapid contraction than in cases where the star remains in thermal equilibrium, as the mass-transfer rate exceeds the thermal adjustment capability of the star. Although the donor still responds on a thermal timescale, this out-of-equilibrium contraction causes the radius to shrink more significantly, allowing the donor to reach the critical radius for full convection earlier \citep{Ge2015,Temmink2023, Lu2025}. Consequently, compared to the Matt12 model, the donor in the Skumanich model enters full convection at an earlier evolutionary stage and with a lower mass.

\citet{Knigge2011} adjusted the MB efficiency factor \( f_{\rm MB} \) to better reproduce the period distribution of CV systems, showing that MB efficiency affects the donor mass at full convection and thus modifies the upper edge of the period gap. \citet{Barraza2025} also noted that stronger MB shortens mass-transfer timescales, causing the donor to become fully convective at slightly lower masses. Furthermore, \citet{larsen2025} developed a simplified model to investigate magnetic field effects on the structure and evolution of low-mass main-sequence donors: using mixing-length theory to describe convective energy transport and simulating magnetic suppression of convection by reducing the mixing-length parameter (\( \alpha \)). As \( \alpha \) decreases, convective efficiency is reduced, inhibiting energy transport and increasing the internal temperature gradient, causing stellar expansion. This leads to longer orbital periods for a given donor mass and lower mass-transfer rate, significantly altering the evolutionary path of the system. Therefore, we conclude that magnetic fields cause the donor to reach full convection earlier (at higher mass), while also explaining the lower AML rate in Matt12 compared to Skumanich.

After MB shuts off, orbital contraction substantially slows down. 
Because the donor had been inflated beyond its thermal-equilibrium radius during the preceding high mass-transfer phase, it enters a phase of thermal relaxation and contracts toward its smaller equilibrium radius. 
This contraction proceeds on the thermal timescale and is faster than the GR-driven orbital shrinkage, causing $R_2$ to decrease more rapidly than the Roche-lobe radius $R_{L,2}$. 
Consequently, the donor shrinks below $R_{L,2}$ and the system enters a detached phase, thereby forming the period gap \citep{stehle1996,Baraffe2000,Howell2001,kolb2005,ge2020}. 
Once detached, mass transfer ceases and mass-loss–related AML vanishes, so the AML during this phase is entirely dominated by GR continues to shrink the orbit until the system reestablishes contact near a 2-hour orbital period.

Figure~\ref{fig:RLOF} compares the evolution of the donor radius and the Roche-lobe radius under different AML prescriptions. For gap models, the donor fills its Roche lobe at $P_{\rm orb}\simeq 4$--5\,h and detaches shortly after MB shuts down near 3\,h, producing a detached interval from $\simeq 3$ to $\simeq 2$\,h. 
In contrast, non-gap models maintain sufficiently high systemic AML---without an abrupt drop near the fully convective boundary---so the donor remains close to Roche-lobe filling throughout the entire short-period regime. 
Continuous mass transfer prevents detachment and results in a remarkably smooth orbital contraction. 
Therefore, in non-gap models, it is the persistence of systemic AML that prevents the formation of the period gap.
All models exhibit a decrease or shutdown of MB, yet they show such different behaviors in $\dot{J}_{\mathrm{ML}}$.
In our calculations, $\dot{J}_{\mathrm{ML}}$ is proportional to the mass-transfer rate ($\dot{J}_{\mathrm{ML}} \propto \dot{M}$), so its evolution directly reflects how strongly the donor contracts when MB ceases. We infer that models with stronger pre-gap MB drive higher mass-transfer rates before the donor becomes fully convective, leaving the donor more inflated and farther from thermal equilibrium at the moment MB shuts off. As a result, gap models undergo more pronounced thermal contraction, causing a rapid drop in $\dot{M}$ and producing a steep decline in $\dot{J}_{\mathrm{ML}}$. 
In contrast, models with weaker pre-gap MB (non-gap models) experience only gradual contraction, and thus show a correspondingly gradual decline in $\dot{J}_{\mathrm{ML}}$.
\begin{figure}[htbp]
  \centering
  \includegraphics[width=0.48\textwidth]{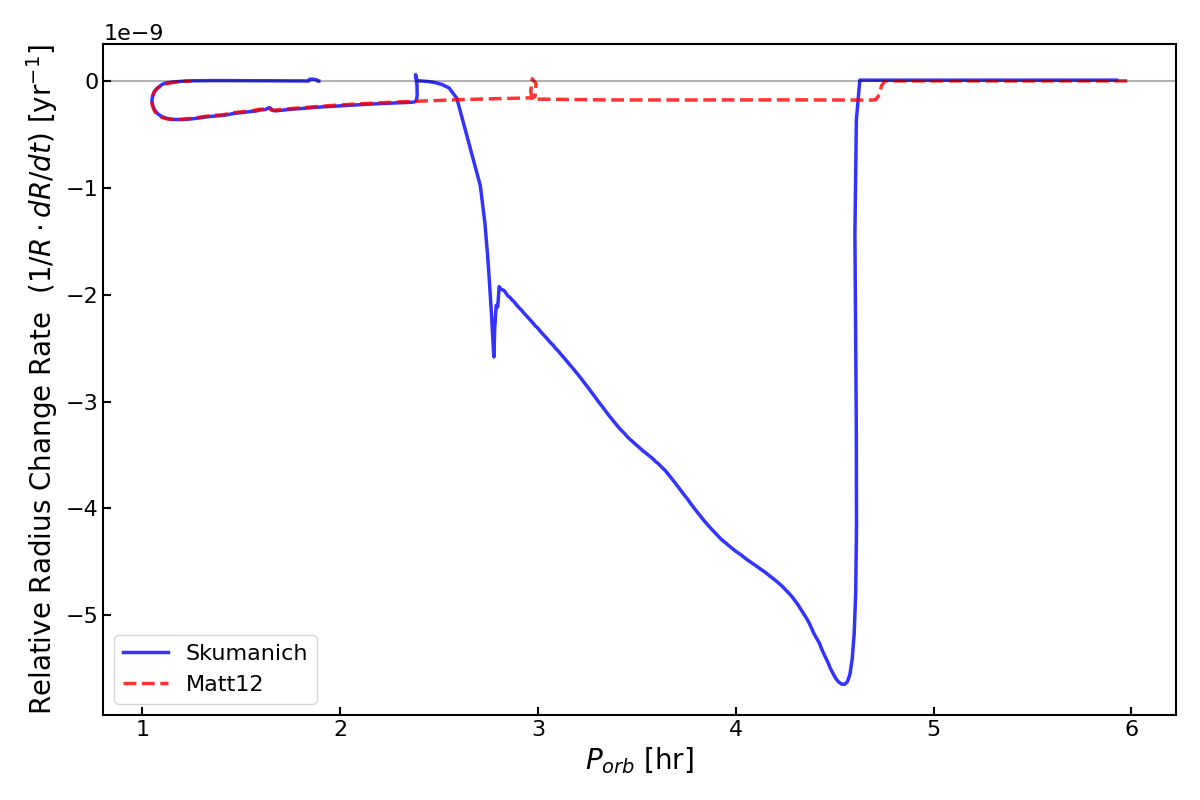}
  \caption{Relative radius change rate $(1/R)\,\mathrm{d}R/\mathrm{d}t$ as a function of orbital period for the Skumanich (solid blue line) and Matt12 (dashed red line) models.}
  \label{fig:radius_change}
\end{figure}
\begin{figure}[htbp]
  \centering
  \includegraphics[width=0.48\textwidth]{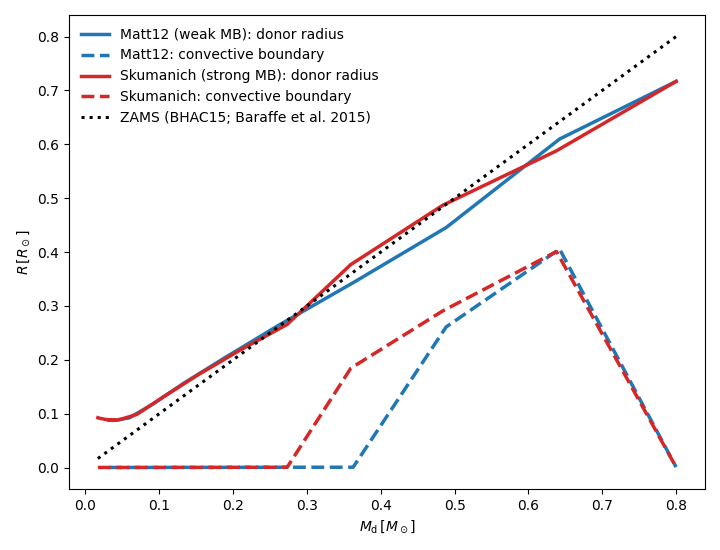}
  \caption{Donor mass–radius relation and convective structure for the Matt12 and Skumanich models. Solid lines show the donor radius, while dashed lines indicate the location of the convective boundary. The dotted line represents the ZAMS mass–radius relation from the BHAC15 stellar models \citep{Baraffe2015}, shown solely as an approximate reference equilibrium sequence.}
  \label{fig:ZAMS}
\end{figure}
\section{Discussion}

Magnetic braking models with different prescriptions exhibit distinct behaviors around the orbital period gap. 
In particular, models that incorporate self-consistent magnetic field regulation parameters generally fail to produce a clear gap.
Although in some magnetic CVs (such as polars and IPs) the donor may possess stronger magnetic fields influenced by the field of the WD—thereby suppressing convective energy transport by altering the mixing-length parameter and regulating the onset of full convection \citep{Feiden2013}—the results of this study indicate that within our model framework, the dominant factor determining whether a system forms a period gap remains the difference in AML strength caused by MB.

For Skumanich-type strong MB models, the large AML drives the donor significantly out of thermal equilibrium. When the donor becomes fully convective at an orbital period of approximately 3 hours, MB ceases abruptly, triggering rapid thermal contraction of the donor and detachment from the Roche lobe, thereby forming the typical 2--3 hour period gap. In contrast, magnetic regulation models such as Matt12 maintain consistently weak AML throughout their evolution.

Figure~\ref{fig:radius_change} shows that the relative radius change rate in the Matt12 model is nearly zero during the short-period phase, indicating the absence of significant thermal readjustment. The donor neither experiences substantial expansion nor undergoes rapid contraction after MB turns off. Figure~\ref{fig:ZAMS} provides a more direct and physically transparent illustration of this behavior by showing the donor mass--radius relation together with the evolution of the convective boundary for the two MB prescriptions, compared to an approximate zero-age main-sequence (ZAMS) reference based on the BHAC15 models \citep{Baraffe2015}. Specifically, during the initial phase of convective development, the donor mass--radius relation in the Matt12 model lies slightly below the ZAMS reference, whereas the Skumanich model is positioned above it. At a given donor mass, the Skumanich model consistently exhibits a somewhat larger donor radius than the Matt12 model.

In the Matt12 model, although the donor does not coincide exactly with the ZAMS mass–radius relation from the BHAC15 stellar models ---which is not expected for a mass-losing star---its slow radius evolution demonstrates the absence of the rapid thermal contraction required to produce Roche-lobe detachment. Consequently, in the no-gap models, the donor radius at MB turn-off remains very close to the expected equilibrium radius, and the donor continues to nearly fill its Roche lobe. As a result, no long-lived detached phase develops and no orbital period gap forms. The fundamental reason for the lack of a clear detached phase therefore lies in the inherently weak AML prior to the cessation of MB.

A recent Letter by \citet{Ort2024} independently investigated CV secular evolution using a MB prescription based on magnetic field complexity, in which increasingly multipolar magnetic field topologies strongly suppress AML. While their specific formulation differs from the MB prescriptions adopted in this work, it belongs to the broader class of self-consistent, self-regulated MB models. \citet{Ort2024} find that such field-complexity-based braking remains too weak to drive the donor significantly out of thermal equilibrium and therefore fails to produce a long-lived detached phase or the canonical orbital period gap. This conclusion is fully consistent with our results for the Matt12 and RM12 prescriptions, which likewise maintain weak pre-gap AML and do not generate a clear period gap. By contrast, our results further demonstrate that a pronounced period gap can be produced only when the donor is driven far from thermal equilibrium by sufficiently strong MB prior to the fully convective boundary.

\begin{figure}[htbp]
  \centering
  \includegraphics[width=0.48\textwidth]{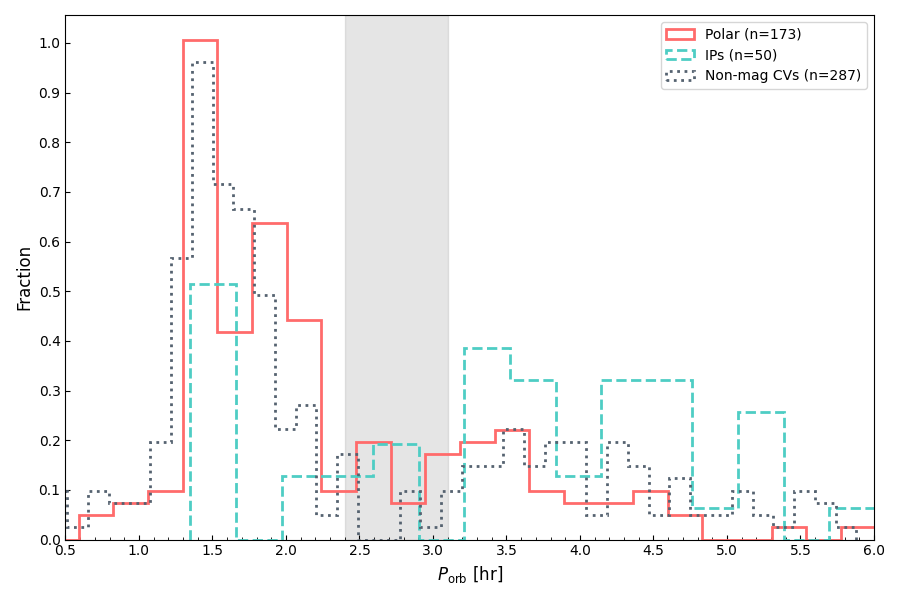}
  \caption{Normalized orbital period distributions for different classes of CVs. Polar systems are represented by solid red lines, IPs by dashed teal lines, and nonmagnetic systems by dotted gray lines. Data were compiled from \citet{Norton2004}, \citet{Ferrario2015}, \citet{Parsons2021}, \citet{Hakala2022}, \citet{van2025}, \citet{Sparks2021}, and \citet{Schreiber2024}. The gray shaded region indicates the period gap proposed by \citet{Schreiber2024}. The numbers shown in the legend (n) denote the number of systems in each subclass.}
  \label{fig:PP}
\end{figure}

Furthermore, we conducted a statistical analysis of the orbital period distributions of magnetic and nonmagnetic CVs (Fig.~\ref{fig:PP}). Given the nature of the literature compilation, the samples drawn from different sources are not fully independent and may include overlapping systems, in particular with the catalog used by \citet{Schreiber2024}. Since Fig.~\ref{fig:PP} shows normalized orbital-period distributions, potential overlaps do not affect the qualitative comparison presented here. The results align with the observational analysis by \citet{Schreiber2024}, which, based on the currently best available CV sample, identifies a well-defined period gap for nonmagnetic CVs between 2.4 and 3.1 hours, while finding no corresponding gap for magnetic systems, thereby providing a more precise observational benchmark for theoretical models. This refined constraint imposes stricter requirements on evolutionary models. While only our strong MB prescriptions successfully produce a detached phase, qualitative differences persist when compared to this observed gap phenomenon. 
Within this refined gap region, magnetic CVs are more numerous than nonmagnetic systems, suggesting they are more common in this orbital period range. 
Since this study primarily adjusts the MB properties of the donor without incorporating the structural evolution of the WD magnetic field, its potential effects on donor structure and AML require further investigation.

\section{Conclusions}
In this study we systematically examined the evolution of CVs under five distinct MB prescriptions: the classical Skumanich model, the Matt12 model, the RM12 model, an intermediate model, and a boosted model. Our analysis reveals that the choice of MB prescription significantly influences the orbital period distribution, mass-transfer rates, and donor star properties.

A key finding of this work is that different MB prescriptions lead to fundamentally distinct evolutionary behaviors around the orbital period gap. Models that can drive stronger AML (e.g., the classical Skumanich model) drive the donor significantly out of thermal equilibrium before it becomes fully convective. When MB ceases abruptly after around 3 hours, rapid thermal contraction causes the donor to detach from the Roche lobe, naturally forming the characteristic 2--3 hour period gap. In contrast, self-consistent magnetic regulation models like Matt12 and RM12 maintain consistently weak AML throughout their evolution, keeping the donor in a quasi-thermal equilibrium state, as demonstrated by its near-equilibrium mass–radius relation shown in Fig.~\ref{fig:ZAMS}. Consequently, when MB terminates, only minimal contraction occurs and the donor continues to fill its Roche lobe, resulting in smooth orbital evolution without a clear detached phase.

Comparison with observed period distributions shows that the intermediate model provides the best overall agreement with observations of nonmagnetic CVs in terms of gap location, width, and donor properties. Our statistical analysis of orbital periods in magnetic and nonmagnetic CVs further supports the observational result reported by \citet{Schreiber2024} that, within the 2--3 hour orbital period range corresponding to the gap in nonmagnetic systems, magnetic CVs are more numerous than nonmagnetic ones.

While the Matt12 and RM12 models fail to reproduce the period gap, they may be more appropriate for explaining magnetic systems such as polars and IPs. However, our current models cannot fully explain these systems since we have not incorporated the structural evolution of the magnetic field of the WD.

Future work should further incorporate the regulatory effects of magnetic fields on donor structure, including convection suppression, radius expansion, and potential influences from  magnetic fields of the WD. Only by more accurately accounting for these effects in models can we hope to faithfully reproduce the period distribution characteristics of CV populations with different magnetic properties.

\begin{acknowledgements}
      This work has been supported by the National Natural Science Foundation of China (Grants 12563007, 12373038, 12288102, 12463011, 12163005, 12003025), the Natural Science Foundation of Xinjiang (Grants 2022TSYCLJ0006, 2022D01D85 and 2024D01C230), the China Manned Space Program (Grant CMS-CSST-2025-A15), the Foundation of Tianshan Talents program 2024TSYCJU0001 and the Major Science and Technology Program of Xinjiang Uygur Autonomous Region under grant 2022A03013-3.
\end{acknowledgements}

\bibliographystyle{aa}
\bibliography{reference}


\appendix
\section{CV data} \label{app:cv_data}

\begin{table}[h]
\caption{CVs with orbital periods and mass-transfer rates.}
\label{tab:cv_new_data}
\centering
\begin{tabular}{l c c c c}
\hline \hline
System name & \multicolumn{1}{c}{Orbital period} & Type & \multicolumn{1}{c}{Mass-transfer rate} & Ref. \\
& \multicolumn{1}{c}{(hr)} & & \multicolumn{1}{c}{($M_{\odot}\,\mathrm{yr}^{-1}$)} & \\
\hline
J0720 & 1.5   & Polar & $2 \times 10^{-12}$ & [1] \\
ZTF J0112+5827 & 1.348 & Polar & $(1.4 - 6.3) \times 10^{-11}$ & [2] \\
IGR J17014-4306 & 12.8 & IP & $2.1 \times 10^{-6}$ & [3] \\
EX Hydrae & 1.638 & IP & $3.9 \times 10^{-11}$ & [4] \\
XMM J1527 & 1.873 & Polar & $(5.0 \pm 1.0) \times 10^{-12}$ & [5] \\
J0154 & 1.48  & Polar & $2.9 \times 10^{-11}$ & [6] \\
J0600 & 1.31  & Polar & $9.7 \times 10^{-11}$ & [6] \\
J0859 & 2.40  & Polar & $6.9 \times 10^{-11}$ & [6] \\
J0953 & 1.73  & Polar & $5.8 \times 10^{-11}$ & [6] \\
J1002 & 1.67  & Polar & $8.7 \times 10^{-11}$ & [6] \\
ZTF J0850 & 1.72  & Polar & $\sim 10^{-11}$ & [7] \\
V379 Vir & 1.473 & Polar & $3.2 \times 10^{-14}$ & [8] \\
SDSS 1514 & 1.478 & Polar & $3.3 \times 10^{-14}$ & [8] \\
SDSS 1250 & 1.438 & Polar & $1.7 \times 10^{-14}$ & [8] \\
\hline
\end{tabular}
\tablebib{(1)~\citet{Bobakov2025};
(2)~\citet{Lin2025};
(3)~\citet{Zang2025};
(4)~\citet{Beuermann2024};
(5)~\citet{Ok2024};
(6)~\citet{Beuermann2021};
(7)~\citet{Rodriguez2023};
(8)~\citet{Munoz2023}.
}
\end{table}

\end{document}